\documentclass[twocolumn,floatfix,aps,prl,superscriptaddress,showpacs]{revtex4-1}
\usepackage[latin9]{inputenc}
\usepackage{bm}
\usepackage{amsmath}
\usepackage{graphicx}
\usepackage[unicode=true,
 bookmarks=false,
 breaklinks=false,pdfborder={0 0 1},colorlinks=false]
 {hyperref}
\hypersetup{
 colorlinks,linkcolor=blue,citecolor=blue,urlcolor=blue}

\makeatletter

\usepackage{dcolumn}
\usepackage{bm}
\usepackage{color}
\usepackage[none]{hyphenat}

\makeatother

\begin{document}
\title{Branched flow of intense laser light in plasma with uneven density distribution}

\author{K. Jiang}
\affiliation{Shenzhen Key Laboratory of Ultraintense Laser and Advanced Material Technology, Center for Advanced Material Diagnostic Technology, and College of Engineering Physics, Shenzhen Technology University, Shenzhen 518118, People's Republic of China}
\affiliation{College of Applied Sciences, Shenzhen University, Shenzhen 518060, People's Republic of China}

\author{T. W. Huang}
\email{taiwu.huang@sztu.edu.cn}
\affiliation{Shenzhen Key Laboratory of Ultraintense Laser and Advanced Material Technology, Center for Advanced Material Diagnostic Technology, and College of Engineering Physics, Shenzhen Technology University, Shenzhen 518118, People's Republic of China}

\author{C. N. Wu}
\affiliation{Graduate School, China Academy of Engineering Physics, Beijing 100088, People's Republic of China}

\author{M. Y. Yu}
\affiliation{Shenzhen Key Laboratory of Ultraintense Laser and Advanced Material Technology, Center for Advanced Material Diagnostic Technology, and College of Engineering Physics, Shenzhen Technology University, Shenzhen 518118, People's Republic of China}

\author{H. Zhang}
\affiliation{Shenzhen Key Laboratory of Ultraintense Laser and Advanced Material Technology, Center for Advanced Material Diagnostic Technology, and College of Engineering Physics, Shenzhen Technology University, Shenzhen 518118, People's Republic of China}

\author{S. Z. Wu}
\affiliation{Shenzhen Key Laboratory of Ultraintense Laser and Advanced Material Technology, Center for Advanced Material Diagnostic Technology, and College of Engineering Physics, Shenzhen Technology University, Shenzhen 518118, People's Republic of China}

\author{H. B. Zhuo}
\affiliation{Shenzhen Key Laboratory of Ultraintense Laser and Advanced Material Technology, Center for Advanced Material Diagnostic Technology, and College of Engineering Physics, Shenzhen Technology University, Shenzhen 518118, People's Republic of China}

\author{A. Pukhov}
\affiliation{Institut f\"{u}r Theoretische Physik I, Heinrich-Heine-Universit\"{a}t D\"{u}sseldorf, 40225 D\"{u}sseldorf, Germany}

\author{C. T. Zhou}
\email{zcangtao@sztu.edu.cn}
\affiliation{Shenzhen Key Laboratory of Ultraintense Laser and Advanced Material Technology, Center for Advanced Material Diagnostic Technology, and College of Engineering Physics, Shenzhen Technology University, Shenzhen 518118, People's Republic of China}
\affiliation{College of Applied Sciences, Shenzhen University, Shenzhen 518060, People's Republic of China}

\author{S. C. Ruan}
\email{scruan@sztu.edu.cn}
\affiliation{Shenzhen Key Laboratory of Ultraintense Laser and Advanced Material Technology, Center for Advanced Material Diagnostic Technology, and College of Engineering Physics, Shenzhen Technology University, Shenzhen 518118, People's Republic of China}
\affiliation{College of Applied Sciences, Shenzhen University, Shenzhen 518060, People's Republic of China}

\date{\today}

\begin{abstract}
Branched flow is an interesting phenomenon that can occur in diverse systems. It is usually linear in the sense that the flow does not alter the medium properties. Branched flow of light on thin films was recently discovered. A question of interest is thus if nonlinear branched flow of light can also occur. Here we found using particle-in-cell simulations that with intense laser propagating in plasma with randomly uneven density distribution, photoionization by the laser can locally enhance the density variations along the laser paths and thus the branching of the laser. However, too-intense lasers can smooth the uneven electron density and suppress branching. The observed branching properties agree well with an analysis based on a Helmholtz equation for the laser electric field. Branched flow of intense laser in uneven plasma potentially opens up a new realm of intense laser-matter interaction.
\end{abstract}
\maketitle
\sloppy{}

Waves propagating in a randomly uneven medium with correlation length larger than the wavelength $\lambda$ can form filaments in a tree-branch-like manner \cite{Heller}. This phenomenon, known as branched flow, has been observed in diverse systems and at many length scales \cite{Topinka,Aidala,Jura,Maryenko,Liu,Berry,Berry1,Kanoglu,Degueldre,Cordes,Pidwerbetsky,Stinebring,Barkhofen,Green,Derr,Patsyk,Patsyk1,Brandst,Mattheakis}. For example, instead of smoothly diffusing or spreading, an electron beam passing through two-dimensional (2D) electron gases can form branching strands that become successively narrower \cite{Topinka,Aidala,Jura,Maryenko,Liu}, giant freak ocean waves are attributed to the random unevenness on the ocean floor \cite{Berry,Berry1,Kanoglu,Degueldre,Green}, and branching of microwave radiation emitted by pulsars is attributed to the interstellar dust clouds \cite{Cordes,Pidwerbetsky,Stinebring}. Recently, branched flow of light (specifically continuous-wave laser) has been found on thin soap films \cite{Patsyk,Patsyk1}. The light branching is attributed to variations of the film's refractivity $\eta$, which bend and bundle the light rays at favorable locations and form caustics \cite{Patsyk,Patsyk1,Brandst,Mattheakis}. Despite its complexity, branched flow is usually a linear phenomenon and the distance $d_0$ from the source to the first branching point follows the scaling $d_0\propto l_cv_0^{-2/3}$ \cite{Heller,Kaplan,Metzger,Metzger1,Zapperi,Barkhofen}, where $l_c$ is the correlation length of the medium's unevenness, and $v_0$ is its strength parameter (to be defined later).

At present, lasers with intensity $I$ ranging from $10^{14}$ to $10^{20}$ W/cm$^2$ are readily available. The propagation of such intense laser through matter sets off new phenomena \cite{Gibbon}. When $I\gtrsim10^{14}$ W/cm$^2$,  the atomic Coulomb barrier is suppressed by the strong laser electric field, electrons are set free and the affected medium is ionized into plasma \cite{Augst}, whose optical properties then become dominated by electron dynamics. Moreover, at higher laser intensities $I\gtrsim1.37\times10^{18}$ W/cm$^2$, namely, the laser intensity is above the relativistic threshold, in addition to photoionization, the laser ponderomotive force and relativistic plasma motion can significantly modify the original unevenness in the density as well as the local refractivity, thereby affecting the laser propagation \cite{Gibbon}. Whether light branching can occur in such intense laser-matter interaction involving with complex nonlinear effects is a timely and critical question.

In this Letter, we present the first investigation of nonlinear branched flow of intense light in laser-matter interaction. Particle-in-cell (PIC) simulations show that laser branching can occur at moderate laser intensities ($I\cong10^{14}-10^{17}$ W/cm$^2$) in inhomogeneous plasma with randomly uneven density distribution. In contrast to linear light branching, in this regime the branching depends on the laser intensity. In particular, photoionization induced by the strong laser electric field raises the density unevenness along the laser paths and enhances the branching. However, relativistic lasers can suppress branching by smoothing the unevenness and thus the local refractivity of the plasma. An analysis of the branching process and the resulting properties consistent with the simulation results is also given.

\begin{figure}
\centering
\includegraphics[width=8.6cm]{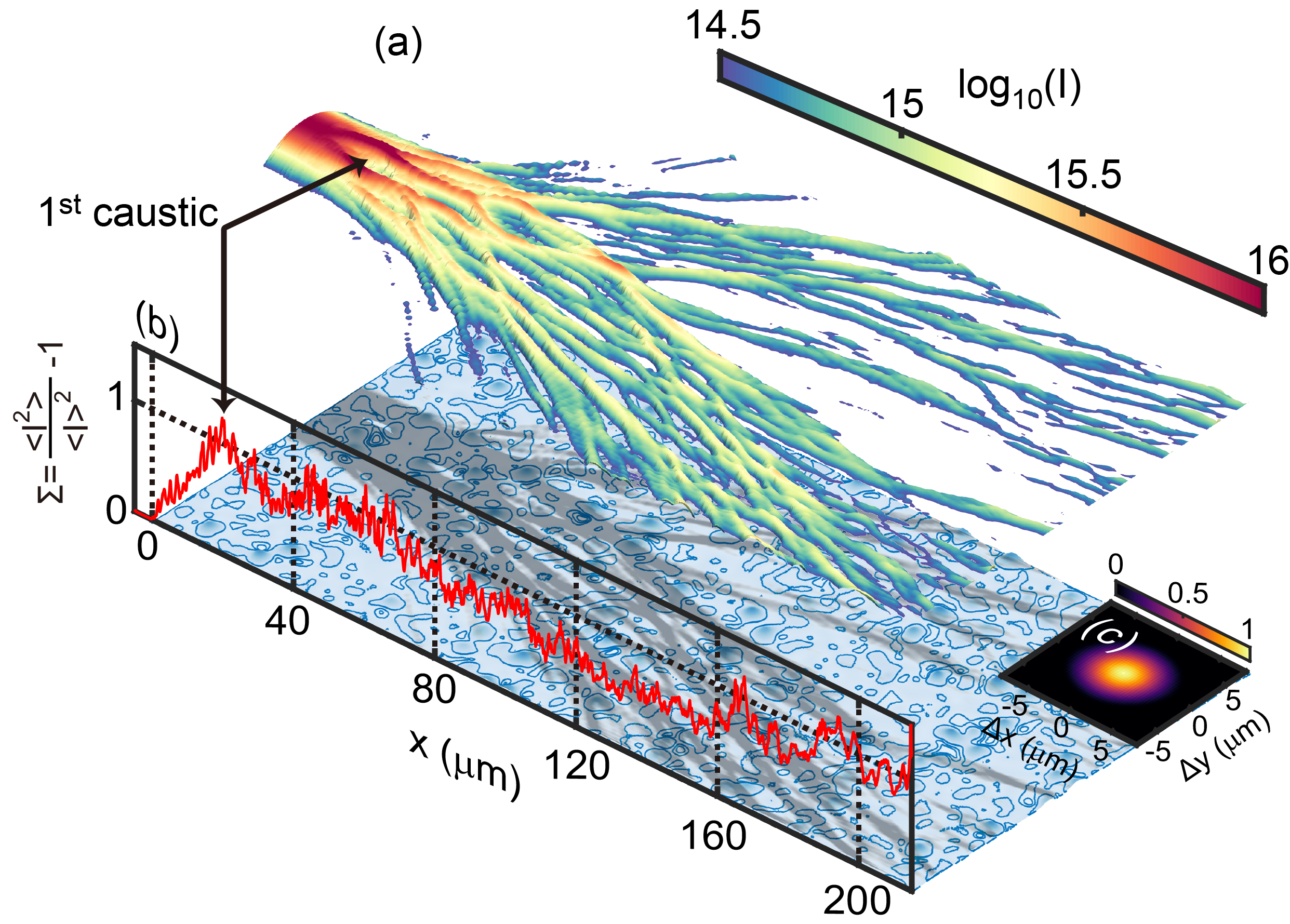}
\caption{(a) Flow branching as an intense laser propagates through an uneven plasma along the $x$ direction. The (logarithmic) color bar is for the laser intensity in W/cm$^2$ at $t=1095$ fs. The initial distribution of electron density is shown (in blue) beneath the laser shadow. For reference, (b) shows the spatial evolution of the scintillation index $\Sigma$ of a planewave with identical interaction parameters propagating through the same plasma with randomly uneven density distribution. The black arrows mark the first caustics, or branching point. The inset (c) shows the autocorrelation function of the initial electron density distribution.}
\label{fig_1}
\end{figure}

Light branching is usually three dimensional. However, if the irregularity of the uneven background medium is isotropic, branching can be effectively 2D \cite{Heller,Kaplan,Klyatskin,Degueldre1,Sup}. Accordingly, we shall conduct 2D PIC simulations of intense-laser branching using the \textsc{epoch} code \cite{Arber}. In the simulations, the initial background medium (see Fig. \ref{fig_1}(a), in blue color at the bottom) is weakly pre-ionized SiO$_2$ plasma (with Si$^{2+}$ and O$^{+}$ ions) with uneven density distribution located in $0<x<215$ $\mu$m, $-55$ $\mu$m$<y<55$ $\mu$m. The density unevenness is of an isotropic correlation length $l_c=4.8$ $\mu$m, as obtained from the autocorrelation function (ACF) \cite{Gadelmawlam} shown in Fig. \ref{fig_1}(c). The average densities of Si$^{2+}$, O$^{+}$, and electrons are $0.02n_c$, $0.04n_c$, and $0.08n_c$, respectively, where $n_c\sim1.1\times10^{21}$ cm$^{-3}/\lambda^2$ is the critical density. The coefficient of density variation is $\sim 30\%$. The simulation box is $-5$ $\mu$m $<x<215$ $\mu$m, $-55$ $\mu$m $<y<55$ $\mu$m, with $2200\times1100$ grid cells and 30 macroparticles per cell for each species. Open lateral boundaries are used. An intense circularly-polarized Gaussian laser pulse of central wavelength $\lambda_0=1.06$ $\mu$m, peak intensity $I_0=10^{16}$ W/cm$^2$, and FWHM spotsize $r_0=16.65$ $\mu$m incidents normally from the left boundary. The laser pulse duration is 1.5 ps, with a flat-top time profile. The effects of photoionization during laser-plasma interaction are self-consistently included in the simulations \cite{Sup}. The uneven plasma can be made by ionizing appropriate chemically fabricated low-density foams with a hundred-picosecond laser \cite{Nagai,Milovich}, similar to that in the GSI electron acceleration experiments using near-critical-density plasma \cite{Rosmej,Rosmej1}, except that their nanosecond ionizing laser pulse is replaced by a hundred-picosecond one \cite{Milovich}. Similar uneven plasma may also be produced by laser interaction with gas clusters \cite{LeeNam}, or in the backward Raman amplification schemes for intense-laser pulse compression \cite{Jia}.

Figure \ref{fig_1}(a) for the distribution of the laser intensity at $t=1095$ fs clearly shows light branching: the laser breaks up into several filaments after the first caustics at $\sim19.8$ $\mu$m. As they propagate, the filaments further break up into narrower and weaker ones in a cascade manner. To our knowledge, such bifurcation of intense laser has not been reported before. A useful quantity for characterizing the branch pattern is the scintillation index $\Sigma=(\langle I^2\rangle/\langle I\rangle^2)-1$, which measures the relative strength of intensity fluctuations \cite{Green}. For statistical accuracy, here $\Sigma$ is obtained from simulations using a reference planewave with otherwise identical interaction parameters. We see that $\Sigma$ increases from 0 at the plasma front surface to 1.23 at $x=19.8$ $\mu$m. The dependence of $\Sigma$ on $x$ agrees well with the branch pattern of the laser intensity. Since optical turbulence is defined by $\Sigma>1$ \cite{Rao}, we can consider that the laser pulse evolves from Gaussian to strongly fluctuating one within only $19.8$ $\mu$m, much shorter than that of typical filamentation instabilities \cite{Shukla,Sun,Borisov,Borisov1,Pukhov,Huang}. We emphasize that the laser branching observed here is quite different from laser filamentation, where a sufficiently intense laser can break up into narrow parallel filaments, usually without further bifurcation as they propagate.

\begin{figure}
\centering
\includegraphics[width=8.6cm]{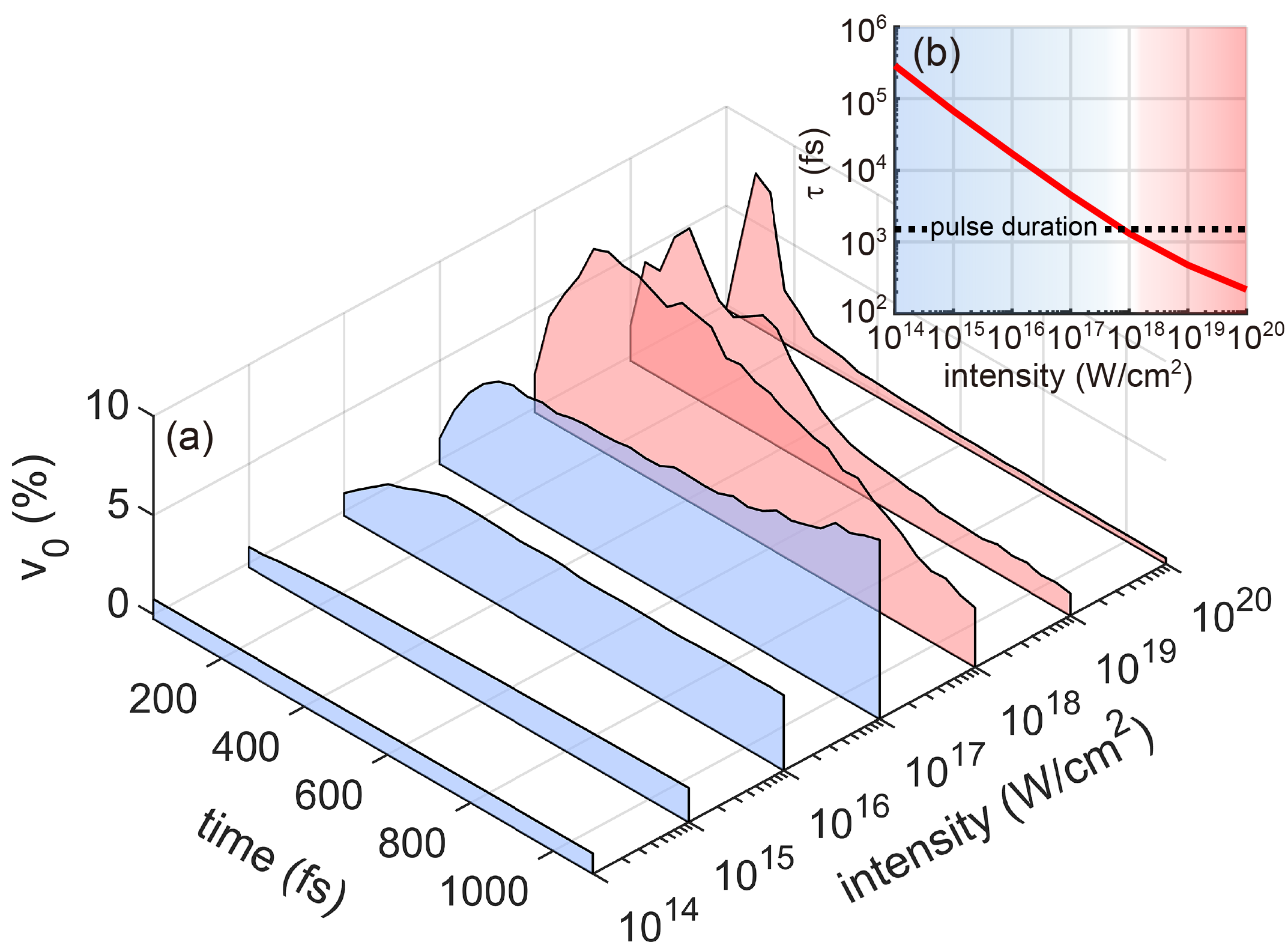}
\caption{(a) Evolution of the potential strength $v_0$ for different laser intensities in the region $0<x<30$ $\mu$m, $-2$ $\mu$m $<y<2$ $\mu$m within the laser spot. (b) Characteristic time $\tau$ of plasma homogenization for different laser intensities. The black dotted line marks the pulse duration. The blue and red patches in (a) and (b) mark the non-relativistic and relativistic intensity regimes, respectively.}
\label{fig_2}
\end{figure}

A feature of intense-laser branching is the nonlinear response of the background plasma medium, which changes the refractivity and the optical unevenness along the laser paths. In Fig. \ref{fig_2}(a), we show the evolution of potential strength $v_0=\sqrt{\left\langle (\bar{\eta^2}-\eta^2_{\mathrm{eff}})^2\right\rangle }/2\bar{\eta^2}$, or unevenness, of the plasma by the action of laser with intensity ranging from $10^{14}$ to $10^{20}$ W/cm$^2$ as obtained from PIC simulations. Here, $\bar{\eta^2}=\left\langle \eta^2_{\mathrm{eff}}\right\rangle $ is the mean-square value of the plasma refractivity $\eta_{\mathrm{eff}}$. For laser intensity below the relativistic threshold ($\lesssim10^{17}$ W/cm$^2$), $v_0$ reaches a plateau after a sharp increase by photoionization. However, in the relativistic regime ($I\gtrsim10^{18}$ W/cm$^2$), $v_0$ after the peaking decrease continuously to even smaller values than the initial one. The decrease is due to plasma homogenization by the laser interaction, at a time scale estimated to be $\tau=l_c/2c_s$, where $c_s=\sqrt{\frac{\left\langle Z^2/m_i\right\rangle T_e}{\left\langle Z\right\rangle }}$ is the ion acoustic speed, $Z$ and $m_i$ are the charge number and rest mass of each ion species, $T_e\sim(\gamma-1)m_ec^2$ is the bulk electron temperature \cite{Wilks}, $\gamma=\sqrt{1+a_0^2}$ for circular polarization, $a_0=e|\bm{E}|/m_e\omega c$ is the normalized laser electric field $\bm{E}$, $m_e$ and $-e$ are the electron rest mass and charge, $\omega$ is the laser frequency, and $c$ is the speed of light in vacuum. One can see in Fig. \ref{fig_2}(b) that for non-relativistic lasers, $\tau$ is much larger than the pulse duration. Therefore, the plasma unevenness $v_0$ can be considered as quasistatic after the rapid increase by photoionization, consistent with the plateaus shown in Fig. \ref{fig_2}(a). In contrast, for relativistic lasers, $\tau$ becomes comparable or smaller than the pulse duration. In this case, removal of the density unevenness by the laser interaction becomes dominant. The background plasma loses its original uneven quality along the laser path, resulting in the pronounced decrease of $v_{0}$. 

For non-relativistic lasers, since the plasma homogenization during the laser interaction can be neglected and the refractivity of the uneven plasma can be considered as slow varying, propagation of the laser can be described by the Helmholtz equation \cite{Patsyk,Sup}
\begin{equation}
-\nabla^2\bm{E}+k_0^2(\bar{\eta^2}-\eta_{\mathrm{eff}}^2)\bm{E}
=k_0^2\bar{\eta^2}\bm{E},\label{eq1}
\end{equation}
where $k_0=2\pi/\lambda_0$ is the wavenumber in vacuum, $\eta_{\mathrm{eff}}=\sqrt{1-n_e(I)/\gamma n_c}$ is the effective refractivity, and $\bar{\eta^2}=\left\langle \eta_{\mathrm{eff}}^2\right\rangle $ with average taken over the laser spot area. Note that the electron number density $n_e$ is explicit function of the instantaneous laser intensity $I$ since electrons are produced by photoionization. In this case, one obtains $v_0=\sqrt{\overline{\delta(I)^2}}/2(n_c-n_e(I)/\gamma)$, where
$\delta(I)=\left\langle n_e(I)/\gamma\right\rangle -n_e(I)/\gamma$ is the local fluctuating strength of the electron density, and $\gamma\sim1$ for non-relativistic lasers. We see that since $v_0$ includes the effect of photoionization, it increases with $I$ due to the increased ionization rate, in agreement with the
simulation results in Fig. \ref{fig_2}(a). Since laser branching is directly related to $v_0$, the
branching is also enhanced by photoionization.

\begin{figure}
\centering
\includegraphics[width=8cm]{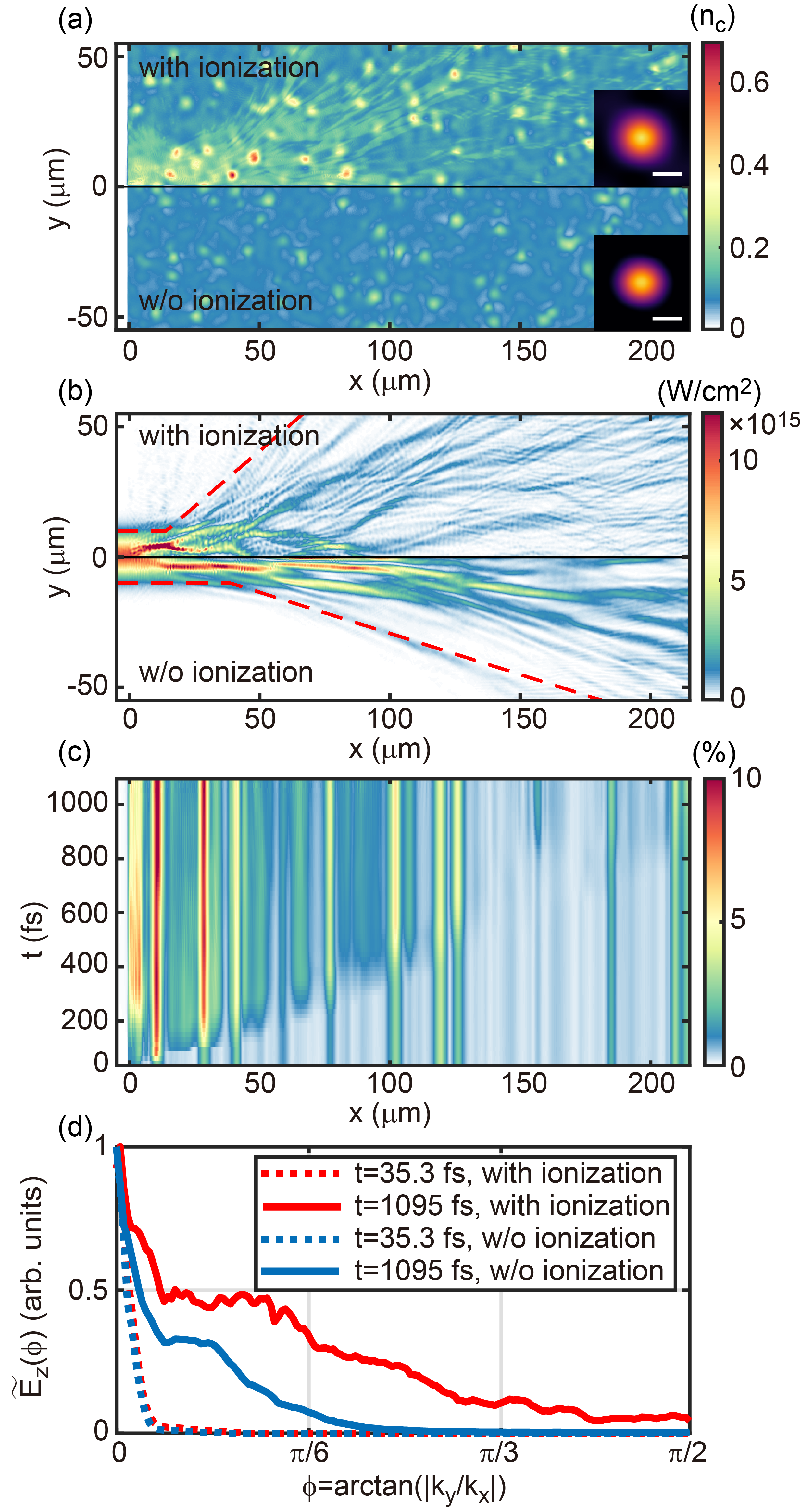}
\caption{Comparison of (a) electron density (in $n_c$) and (b) laser intensity (in W/cm$^2$) at
$t=1095$ fs for with/without photoionization. (c) Evolution of the local potential strength
$v_0(x)$. (d) The strength $\tilde{E}_z(\phi)$ (see the text for definition) at $t=35.3$ and $1095$
fs for with/without photoionization. The insets in (a) show the autocorrelation functions of the
electron density, where the white bars are of length $5$ $\mu$m. The red dashed lines in (b) show the
spread angle of the laser branches.} \label{fig_3}
\end{figure}

Comparison with simulations where photoionization is switched off further confirms the above analysis. As shown Fig. \ref{fig_3}(a), considerable increase of the electron density $n_e$ along the laser paths is observed if photoionization is included. At the laser intensity $I=1\times10^{16}$ W/cm$^2$, tunnelling ionization of both the Si and O ions to the $+4$ state can occur \cite{Augst,Sup}. For our simulation parameters, the average $n_e$ around the laser axis is $\sim0.24n_c$, three-fold of the initial value. In addition, since plasma homogenization can be ignored in the non-relativistic regime, the electrons have the initial disordered distribution of the ions \cite{Sup}. Therefore, the ACF of the electron density at $t=1095$ fs remains almost the same as the initial one. Likewise, the local potential strength $v_0(x)$ keeps its initial correlations (indicated by the vertical streaks) after the rapid increase due to photoionization, as shown in Fig. \ref{fig_3}(c). Note that the correlation length $l_c$ of 4.8 $\mu$m is much smaller than the laser spotsize, i.e. the effect of Gaussian intensity profile across the beam on $l_c$ can be ignored. The increase of $v_0$ leads to stronger (but still relatively weak) scattering of the laser, resulting in the enhancement of the branch pattern, as shown in Fig. \ref{fig_3}(b).

To further characterize the branching, we consider the angle dependence of the laser electric field in the Fourier space, defined by $\tilde{E}_z(\phi)=|\int e^{-ik_0x\cos\phi}e^{-ik_0y\sin\phi}E_z(\phi)\mathrm{d}x\mathrm{d}y|$, where $\phi=\arctan(|k_y/k_x|)$. Here, $E_z$ is used instead of $E_y$ to exclude the self-generated fields. As shown in Fig. \ref{fig_3}(d), the dependence of $\tilde{E}_z(\phi)$ on $\phi$ at $t=35.3$ fs is quite similar for both cases, indicating that most of the laser energy still flows in the $x$ direction. However, at $t=1095$ fs, as a result of many successive weak scatterings in the uneven plasma, a large amount of laser light is branched into other directions. The spread angle $\Theta$ of the light branches after the first caustics can be defined as that when $\tilde{E}_z$ drops to $1/4$ of its maximum. We find $\Theta\sim2\pi/9$ when photoionization is included,  which is about two times larger than that without photoionization. This result further demonstrates that photoionization enhances the unevenness of the refractivity and thus the branching.

\begin{figure}
\centering
\includegraphics[width=8cm]{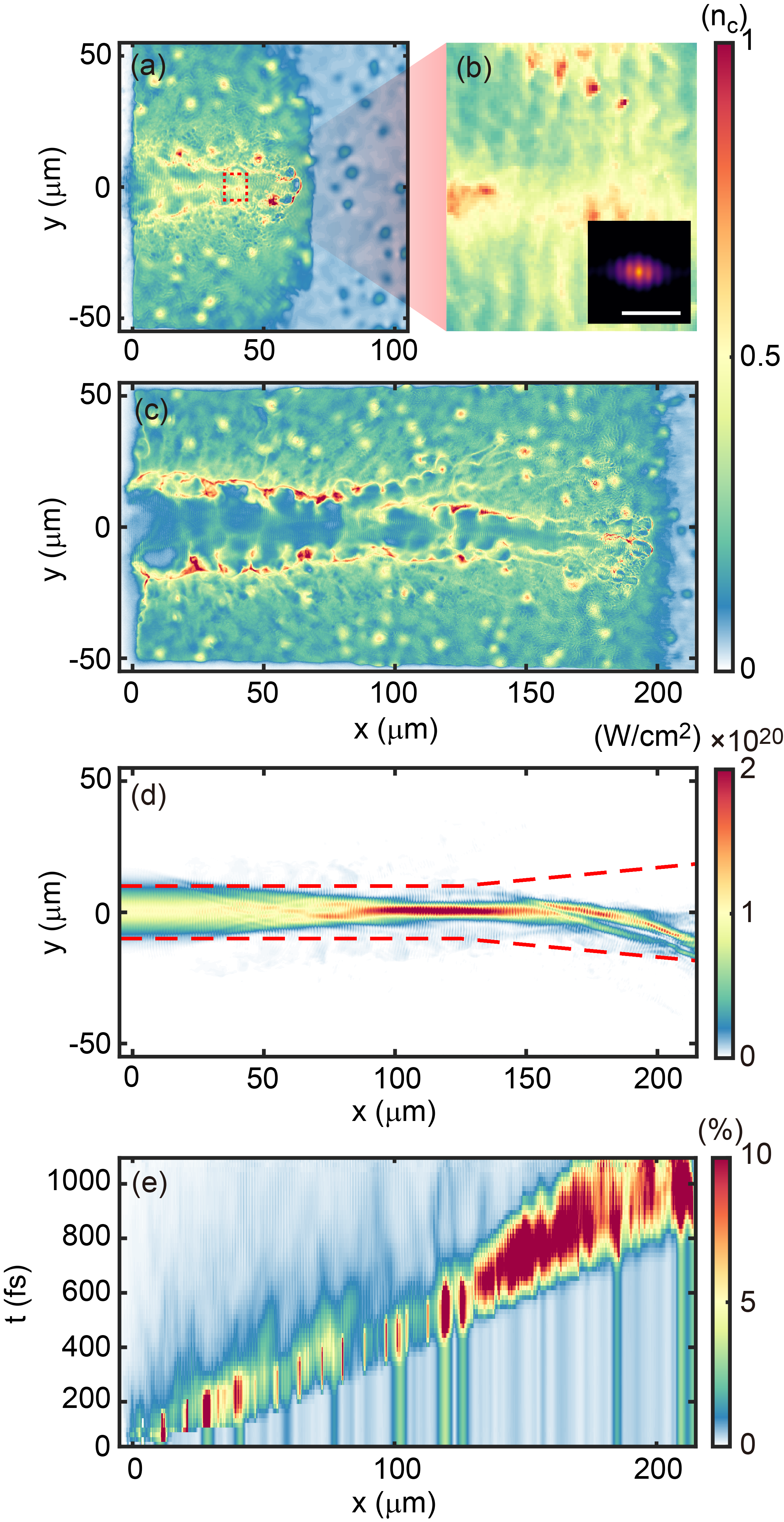}
\caption{(a) Distribution of the electron density (in $n_c$) at $t=283$ fs. (b) Enlargement of the region in the red dashed square in (a). The inset shows the autocorrelation functions of the electron density in this region, and the white bar is of length $5$ $\mu$m. (c) Distribution of electron density at $t=1095$ fs. (d) Distribution of laser intensity (in W/cm$^2$) at $t=1095$ fs. (e) Evolution of the local potential strength $v_0(x)$. The red dashed lines in (d) mark the spread angle of the laser light.}
\label{fig_4}
\end{figure}

For $I>1.37\times10^{18}$ W/cm$^2$ relativistic laser, most of the electrons on the outer shells of the ions are freed and they can be accelerated to light speed by the laser fields within a single cycle. In this case, further ionization becomes marginal and relativistic laser-plasma-interaction effects become significant. The plasma homogenization time $\tau \sim 217$ fs becomes much smaller than the pulse duration. The local refractivity along the laser path now changes simultaneously as the laser propagates, and Eq. \eqref{eq1} becomes inapplicable \cite{Sup}. In fact, Figs. \ref{fig_4}(a) and (b) show that the unevenness in the initial electron density distribution vanishes right behind the laser pulse front. Rapid plasma homogenization leads to the decrease of $l_c$, and electron resonance in the laser fields causes longitudinal modification of the density distribution and the ACF, as can be seen in Fig. \ref{fig_4}(b). In addition, the strong laser ponderomotive force expels the affected electrons, resulting in the formation of plasma channel behind the laser front  \cite{Sun,Borisov,Borisov1,Pukhov}, and further reduction of the density unevenness, as shown in Fig. \ref{fig_4}(c). Figure \ref{fig_4}(e) shows that the corresponding local potential strength $v_0(x)$ decreases to much less than the initial one after the rapid increase caused by photoionization. The initial correlation of the unevenness also vanishes. As shown in Fig. \ref{fig_4}(d), branching of the laser is suppressed and its spread angle $\Theta$ remains small at $2\pi/67$.

\begin{figure}
\centering
\includegraphics[width=8.6cm]{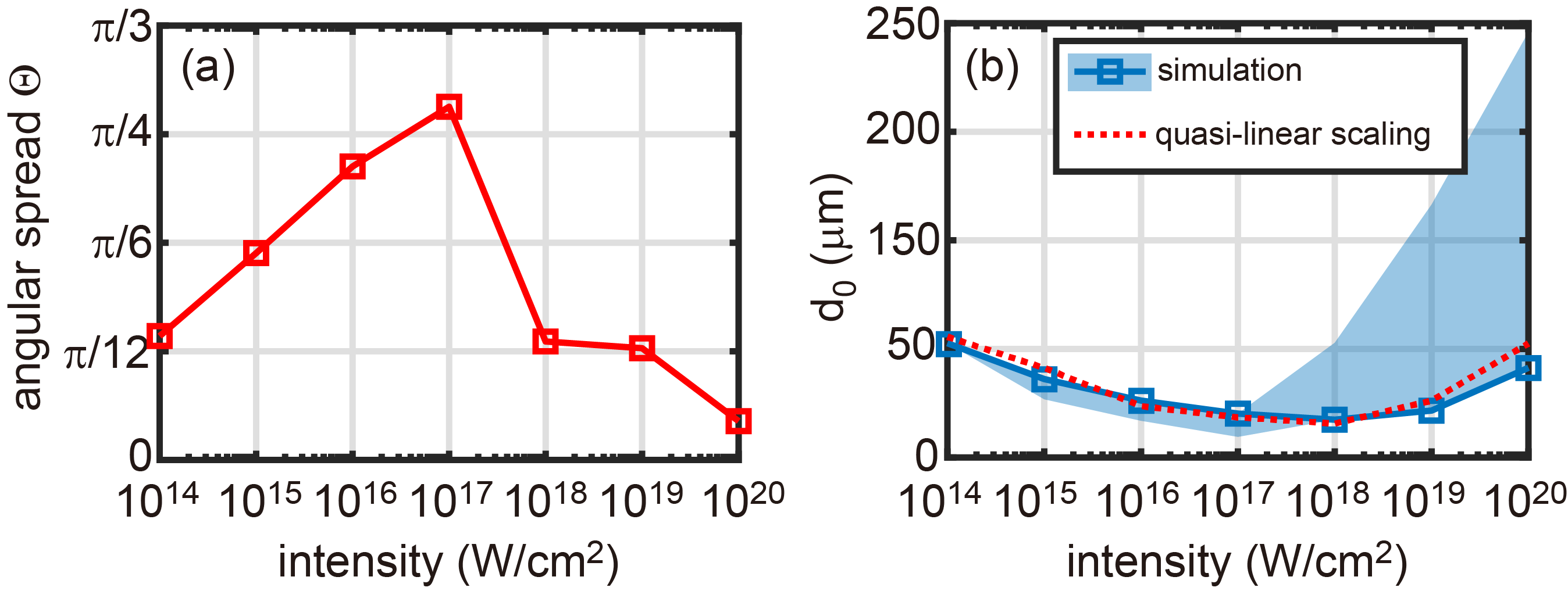}
\caption{(a) The spread angle $\Theta$ (in radians) of the laser branches at $t=1095$ fs for different initial laser intensities. (b) The distance from the source to the first caustics $d_0$ (in $\mu$m) for different laser intensities. The blue solid curve is obtained from simulations at the time when the first caustics appear. The blue patch shows the evolution of $d_0$. The red dashed curve is the result from quasi-linear scaling.}
\label{fig_5}
\end{figure}

Figure \ref{fig_5}(a) for the spread angle $\Theta$ of the laser branches for different initial laser intensities shows that $\Theta$ increases with $I_0$ until $I_0\lesssim10^{17}$ W/cm$^2$, then it decreases as $I_0$ increases further. This is in good agreement with the dependence of the potential strength $v_0$ on the laser intensity at $t=1095$ fs shown in Fig. \ref{fig_2}(a).
Such dependence of $\Theta$ on $I_0$ can be considered as evidence of nonlinear branched flow of intense laser light in experiments. 

Another parameter for characterizing flow-branching is the distance $d_0$ from the boundary (where the flow enters) of the uneven medium to the first branching point. In the linear case, the flow has no influence on the medium, and a universal scaling law for $d_0$ is $d_0\propto l_cv_0^{-2/3}$. For non-relativistic picosecond lasers where plasma homogenization can be ignored, and $v_0=\sqrt{\overline{\delta(I)^2}}/2(n_c-n_e(I)/\gamma)$, one can obtain
\begin{equation}
d_0\propto l_c\left\langle \delta(I)^2\right\rangle ^{-1/3}\left\langle n_c-\frac{n_e(I)}{\gamma}\right\rangle ^{2/3}.\label{eq2}
\end{equation}
Note that the laser intensity appears in the scaling, which is due to the nonlinear effect of photoionization. We see that $d_0$ decreases with increase of the density fluctuating strength given by $\left\langle \delta(I)^2\right\rangle $, and increases with $\left\langle n_c-n_e(I)/\gamma\right\rangle $, indicating that the higher the effective plasma density $n_e/\gamma$ is, the earlier branching occurs. The quasilinear scaling agrees well with the branched flow of non-relativistic lasers, as shown in Fig. \ref{fig_5}(b). However, for $I>10^{18}$ W/cm$^2$ relativistic lasers, laser branching becomes suppressed due to plasma homogenization. The first caustics, instead of being a branching point, now mark the location where self-focusing starts. Figure \ref{fig_5}(b) shows that $d_0$ now increases with the laser intensity. It is of interest to note that Eq. \eqref{eq2} still agrees fairly well with the simulation results, even though Eq. \eqref{eq1} no longer holds in this regime.

In conclusion, we have shown that an intense laser propagating through uneven plasma can form complex light branches. Photoionization can raise unevenness in the density, and thus enhance branch formation. However, relativistic effects of too-intense lasers can suppress branch formation by smoothing the plasma unevenness. These regimes can be potentially verified by experiments based on laser interaction with pre-ionized low-density fibrous or foamy materials, or gas clusters. Our work extends the existing studies of optical branching to the nonlinear regime. The results can be relevant to optical communications, nonlinear optics, strong field physics, as well as laser interaction with foam or turbulent plasma.

This work is supported by the National Key R\&D Program of China (Grant No. 2016YFA0401100), the National Natural Science Foundation of China (Grants No. 12175154, No. 11875092, and No. 12005149), the Natural Science Foundation of Top Talent of SZTU (Grant No. 2019010801001 and 2019020801001), and GCS Juelich (project QED20) in Germany. The EPOCH code is used under UK EPSRC contract (EP/G055165/1 and EP/G056803/1). K. J. would like to thank X. Luo, X. F. Shen, H. Peng, and T. Y. Long for useful discussions.


\section*{References}
\begin{thebibliography}{10}
\bibitem{Heller} E. J. Heller, R. Fleischmann, and T. Kramer, Branched flow. Phys. Today \textbf{74}, 44 (2021).

\bibitem{Topinka} M. A. Topinka, B. J. LeRoy, R. M. Westervelt, S. E. J. Shaw, R. Fleischmann, E. J. Heller, K. D. Maranowski, and A. C. Gossard, Coherent branched flow in a two-dimensional electron gas.  Nature (London) \textbf{410}, 183 (2001).

\bibitem{Aidala} K. E. Aidala, R. E. Parrott, T. Kramer, E. J. Heller,  R. M. Westervelt, M. P. Hanson, and A. C. Gossard, Imaging magnetic focusing of coherent electron waves. Nat. Phys. \textbf{3}, 464 (2007).

\bibitem{Jura} M. P. Jura, M. A. Topinka, L. Urban, A. Yazdani, H. Shtrikman, L. N. Pfeiffer, K. W. West, and D. Goldhaber-Gordon, Unexpected features of branched flow through high-mobility two-dimensional electron gases. Nat. Phys. \textbf{3}, 841 (2007).

\bibitem{Maryenko} D. Maryenko, F. Ospald, K. von Klitzing, J. Smet, J. J. Metzger, R. Fleischmann, T. Geisel, and V. Umansky, How branching can change the conductance of ballistic semiconductor devices. Phys. Rev. B \textbf{85}, 195329 (2012).

\bibitem{Liu} B. Liu and E. J. Heller, Stability of branched flow from a quantum point contact. Phys. Rev. Lett. \textbf{111}, 236804 (2013).

\bibitem{Berry} M. V. Berry, Tsunami asymptotics. New J. Phys. \textbf{7}, 129 (2005).

\bibitem{Berry1} M. V. Berry, Focused tsunami waves. Proc. Roy. Soc. A \textbf{463}, 3055 (2007).

\bibitem{Kanoglu} U. Kanoglu, V. V. Titov, B. Aydin, C. Moore, T. S. Stefanakis, H. Q. Zhou, M. Spillane, and C. E. Synolakis, Focusing of long waves with finite crest over constant depth. Proc. R. Soc. A \textbf{469}, 20130015 (2013).

\bibitem{Degueldre} H. Degueldre, J. J. Metzger, T. Geisel, and R. Fleischmann, Random focusing of tsunami waves. Nat. Phys. \textbf{12} 259 (2016).

\bibitem{Green} G. Green and R. Fleischmann, Branched flow and caustics in nonlinear waves. New J. Phys. \textbf{21}, 083020 (2019).

\bibitem{Cordes} J. Cordes, A. Pidwerbetsky, and R. Lovelace, Refractive and diffractive scattering in the interstellar medium. Astrophys. J. \textbf{310} 737 (1986).

\bibitem{Pidwerbetsky} A. Pidwerbetsky, Simulation and analysis of wave propagation through random media. Ph.D. thesis, Cornell University (1988).

\bibitem{Stinebring} D. R. Stinebring, Scintillation Arcs: Probing Turbulence and Structure in the ISM. Chin. J. Astron. Astrophys. \textbf{6}, 204 (2006).

\bibitem{Barkhofen} S. Barkhofen, J. J. Metzger, R. Fleischmann, U. Kuhl, and H.-J. St${\rm {\ddot{o}}}$ckmann, Experimental Observation of a fundamental length scale of waves in random media. Phys. Rev. Lett. \textbf{111}, 183902 (2013).

\bibitem{Derr} N. J. Derr, D. C. Fronk, C. A. Weber, A. Mahadevan, C. H. Rycroft, and L. Mahadevan, Flow-driven branching in a frangible porous medium. Phys. Rev. Lett. \textbf{125}, 158002 (2020).

\bibitem{Patsyk} A. Patsyk, U. Sivan, M. Segev, and M. A. Bandres, Observation of branched flow of light. Nature \textbf{583}, 60 (2020).

\bibitem{Patsyk1} A. Patsyk, Y. Sharabi, U. Sivan, and M. Segev, Incoherent Branched Flow of Light. Phys. Rev. X \textbf{12}, 021007 (2022).

\bibitem{Brandst} A. Brandst${\rm {\ddot{o}}}$ttera, A. Girschika, P. Ambichla, and S. Rottera, Shaping the branched flow of light through disordered media. Proc. Natl. Acad. Sci. \textbf{116}, 13260 (2019).

\bibitem{Mattheakis} M. Mattheakis and G. P. Tsironis, Extreme waves and branching flows in optical media. Quodons in Mica. 425 (2015).

\bibitem{Kaplan} L. Kaplan, Statistics of branched flow in a weak correlated random potential. Phys. Rev. Lett. \textbf{89}, 184103 (2002).

\bibitem{Metzger} J. J. Metzger, R. Fleischmann, and T. Geisel, Universal statistics of branched flows. Phys. Rev. Lett. \textbf{105}, 020601 (2010).

\bibitem{Metzger1} J. J. Metzger, R. Fleischmann, and T. Geisel, Statistics of extreme waves in random media. Phys. Rev. Lett. \textbf{112}, 203903 (2014).

\bibitem{Zapperi} S. Zapperi, P. Ray, H. E. Stanley, and A. Vespignani, First-order transition in the breakdown of disordered media. Phys. Rev. Lett. \textbf{78}, 1408 (1997).

\bibitem{Gibbon} P. Gibbon. Short pulse laser interactions with matter: an introduction. (World Scientific, 2005).

\bibitem{Augst} S. Augst, D. Strickland, D. D. Meyerhofer, S. L. Chin, and J. H. Eberly, Tunneling ionization of noble gases in a high-intensity laser field. Phys. Rev. Lett. \textbf{63}, 2212 (1989).

\bibitem{Palaniyappan} S. Palaniyappan, 
B. M. Hegelich, H.-C. Wu, D. Jung, D. C. Gautier, L. Yin, B. J. Albright, R. P. Johnson, T. Shimada, S. Letzring, D. T. Offermann, J. Ren, C. K. Huang, R. H\"orlein, B. Dromey, J. C. Fernandez, and R. C. Shah, Dynamics of relativistic transparency and optical shuttering in expanding overdense plasmas. Nat. Phys. \textbf{8}, 763 (2012).

\bibitem{Klyatskin} V. I. Klyatskin, Caustics in random media. Waves in Random Media, \textbf{3}, 93 (1993)

\bibitem{Degueldre1} H. Degueldre, J. J. Metzger, E. Schultheis, and R. Fleischmann, Channeling of branched flow in weakly scattering anisotropic media. Phys. Rev. Lett. \textbf{118}, 024301 (2017).

\bibitem{Sup} See Supplementary Materials for the discussions on particle-in-cell simulations incorporated with photoionization models, ion dynamics, validity of using Helmholtz equation, and dimensional effects, which includes Refs. [1, 21, 26, 28, 29, 31].


\bibitem{Arber} T. D. Arber, K. Bennett, C. S. Brady, A. Lawrence-Douglas, M. G. Ramsay, N. J. Sircombe, P. Gillies, R. G. Evans, H. Schmitz, A. R. Bell, and C. P. Ridgers, Contemporary particle-in-cell approach to laser-plasma modelling. Plasma Phys. Control. Fusion \textbf{57}, 113001 (2015).

\bibitem{Gadelmawlam} E. S. Gadelmawlam, M. M. Koura, T. M. A. Maksoud, I. M. Elewa, and H. H. Soliman, Roughness parameters. J. Mater. Sci. \textbf{123}, 133 (2002).

\bibitem{Nagai} K. Nagai,  C. S. A. Musgrave, and W. Nazarov, A review of low density porous materials used in laser plasma experiments. Phys. Plasmas \textbf{25}, 030501 (2018).

\bibitem{Milovich} J. L. Milovich, O. S. Jones, R. L. Berger, G. E. Kemp, J. S. Oakdale, J. Biener, M. A. Belyaev, D. A. Mariscal, S. Langer, P. A. Sterne, S. Sepke, and M. Stadermann, Simulation studies of the interaction of laser radiation with additively manufactured foams. Plasma Phys. Control. Fusion \textbf{63} 055009 (2021).

\bibitem{Rosmej} O. N. Rosmej, N. E. Andreev, S. Z\"{a}hter, N. Zahn, P. Christ, B. Borm, T. Radon, A. Sokolov, L. P. Pugachev, D. Khaghani, F. Horst, N. G. Borisenko, G. Sklizkov, and V. G. Pimenov, Interaction of relativistically intense laser pulses with long-scale near critical plasmas for optimization of laser based sources of MeV electrons and gamma-rays. New J. Phys. \textbf{21} 043044 (2019).

\bibitem{Rosmej1} O. N. Rosmej, M. Gyrdymov, M. M. G\"{u}nther, N. E. Andreev, P. Tavana, P. Neucaner, S. Z\"{a}hter, N. Zahn, V. S. Popov, N. G. Borisenko, A. Kantsyrev, A. Skobliakov, V. Panyushkin, A. Bogdanov, F. Consoli, X. F. Shen, and A. Pukhov, High-current laser-driven beams of relativistic electrons for high energy density research. Plasma Phys. Control. Fusion \textbf{62} 115024 (2020).

\bibitem{LeeNam} B. R. Lee, P. K. Singh, Y. J. Rhee, C. H. Nam, Spatiotemporal characteristics of high-density gas jet and absolute determination of size and density of gas clusters. Sci. Rep. \textbf{10}, 12973 (2020).

\bibitem{Jia} Q. Jia, K. N. Qu, and N. J. Fisch, Optical phase conjugation in backward Raman amplification. Opt. Lett. \textbf{45}, 5254 (2020).

\bibitem{Rao} R. Z. Rao, Scintillation index of optical wave propagating in turbulent atmosphere. Chin. Phys. B, \textbf{18}, 581 (2009).

\bibitem{Shukla} P. K. Shukla, N. N. Rao, M. Y. Yu, and N. L. Tsintsadze, Relativistic nonlinear effects in plasmas. Phys. Reports \textbf{138}, 1 (1986)

\bibitem{Sun} G. Z. Sun, E. Ott, Y. C. Lee, and P. Guzdar, Self-focusing of short intense pulses in plasmas. Phys. Fluids \textbf{30}, 526 (1987);

\bibitem{Borisov} A. B. Borisov, A. V. Borovskiy, O. B. Shiryaev, V. V. Korobkin, A. M. Prokhorov, J. C. Solem, T. S. Luk, K. Boyer, and C. K. Rhodes, Relativistic and charge-displacement self-channeling of intense ultrashort laser pulses in plasmas. Phys. Rev. A \textbf{45}, 5830 (1992);

\bibitem{Borisov1} A. B. Borisov, A. V. Borovskiy, A. Mcpherson, K. Boyer, and C. K. Rhodes, Stability analysis of relativistic and charge-displacement self-channelling of intense laser pulses in underdense plasmas. Plasma Phys. Control. Fusion \textbf{37}, 569 (1995).

\bibitem{Pukhov} A. Pukhov and J. Meyer-ter-Vehn, Relativistic magnetic self-channeling of light in near-critical plasma: three-dimensional particle-in-cell simulation. Phys. Rev. Lett. \textbf{76}, 3975 (1996).

\bibitem{Huang} T. W. Huang, C. T. Zhou, A. P. L. Robinson, B. Qiao, H. Zhang, S. Z. Wu, H. B. Zhuo, P. A. Norreys, and X. T. He, Mitigating the relativistic laser beam filamentation via an elliptical beam profile. Phys. Rev. E \textbf{92}, 053106 (2015).

\bibitem{Wilks} S. C. Wilks, W. L. Kruer, M. Tabak, and A. B. Langdon, Absorption of ultra-intense laser pulses. Phys. Rev. Lett. \textbf{69}, 1383 (1992).

\end{thebibliography}
\end{document}